\documentclass{article}
\usepackage{ijcai22}

\usepackage{times}
\usepackage{soul}
\usepackage{url}
\usepackage[hidelinks]{hyperref}
\usepackage[utf8]{inputenc}
\usepackage[small]{caption}
\usepackage{graphicx}
\usepackage{amsmath}
\usepackage{amsthm}
\usepackage{booktabs}
\usepackage{algorithm}
\usepackage{algorithmic}
\usepackage{threeparttable, multirow,booktabs,bbding,amsmath,makecell,color}

\title{MNet: Rethinking 2D/3D Networks for Anisotropic Medical Image Segmentation}

\author{
Zhangfu Dong$^{1}$\and
Yuting He$^1$\and
Xiaoming Qi$^1$\and
Yang Chen$^{1,2,3}$\and
Huazhong Shu$^{1,2,3}$\and \\
Jean-Louis Coatrieux$^{2,3}$\and
Guanyu Yang$^{1,2,3,}$\footnote{Corresponding Author}\And
Shuo Li$^4$
\affiliations
$^1$LIST, Key Laboratory of Computer Network and Information Integration (Southeast University), Ministry of Education, Nanjing, China\\
$^2$Jiangsu Provincial Joint International Research Laboratory of Medical Information Processing\\
$^3$Centre de Recherche en Information Biom\'{e}dicale Sino-Fran\c{c}ais (CRIBs)\\
$^4$Dept. of Medical Biophysics, University of Western Ontario, London, ON, Canada
\emails
yang.list@seu.edu.cn
}

\begin{document}

\maketitle

\begin{abstract}
The nature of thick-slice scanning causes severe inter-slice discontinuities of 3D medical images, and the vanilla 2D/3D convolutional neural networks (CNNs) fail to represent sparse inter-slice information and dense intra-slice information in a balanced way, leading to severe underfitting to inter-slice features (for vanilla 2D CNNs) and overfitting to noise from long-range slices (for vanilla 3D CNNs). In this work, a novel mesh network (MNet) is proposed to balance the spatial representation inter axes via learning. 1) Our MNet latently fuses plenty of representation processes by embedding multi-dimensional convolutions deeply into basic modules, making the selections of representation processes flexible, thus balancing representation for sparse inter-slice information and dense intra-slice information adaptively. 2) Our MNet latently fuses multi-dimensional features inside each basic module, simultaneously taking the advantages of 2D (high segmentation accuracy of the easily recognized regions in 2D view) and 3D (high smoothness of 3D organ contour) representations, thus obtaining more accurate modeling for target regions. Comprehensive experiments are performed on four public datasets (CT\&MR), the results consistently demonstrate the proposed MNet outperforms the other methods. The code and datasets are available at: https://github.com/zfdong-code/MNet
\end{abstract}

\section{Introduction}
The rise of deep learning greatly drives the advance in 3D medical image segmentation \cite{lei2020medical}, however, it is still extremely challenging in the 3D images with anisotropic voxel spacing \cite{tajbakhsh2020embracing}. The nature of thick-slice scanning \cite{goldman2007principles} (e.g. CT, MR) causes severe inter-slice discontinuities of 3D medical images, and the vanilla 2D/3D networks \cite{zhou2019unet++,huang2017densely} fail to represent sparse inter-slice information and dense intra-slice information in a balanced way. Specifically, \textbf{1)} underfitting to inter-slice correlations: 2D CNNs \cite{zhou2019unet++,zotti2018convolutional} concentrate on the representation of dense intra-slice information, but completely fail to capture inter-slice correlations, making severe underfitting to inter-slice features, resulting in relatively stable but weak performance (Figure \ref{robust} in experiments section). \textbf{2)} Overfitting to inter-slice noise: 3D CNNs \cite{jiang2021ala,milletari2016v} simultaneously perform convolution on $x/y$- and $z$-axis without taking the anisotropic nature into account. The highly discontinuous sparse information along $z$-axis adds substantial  long-range  noise to the representation of local features in $xy$-plane, making the network prone to over-fitting, resulting in poor generalization ability with the aggravation of anisotropic degree (Figure \ref{robust}).

\begin{figure*}[!htb]
    \centering
    \includegraphics[]{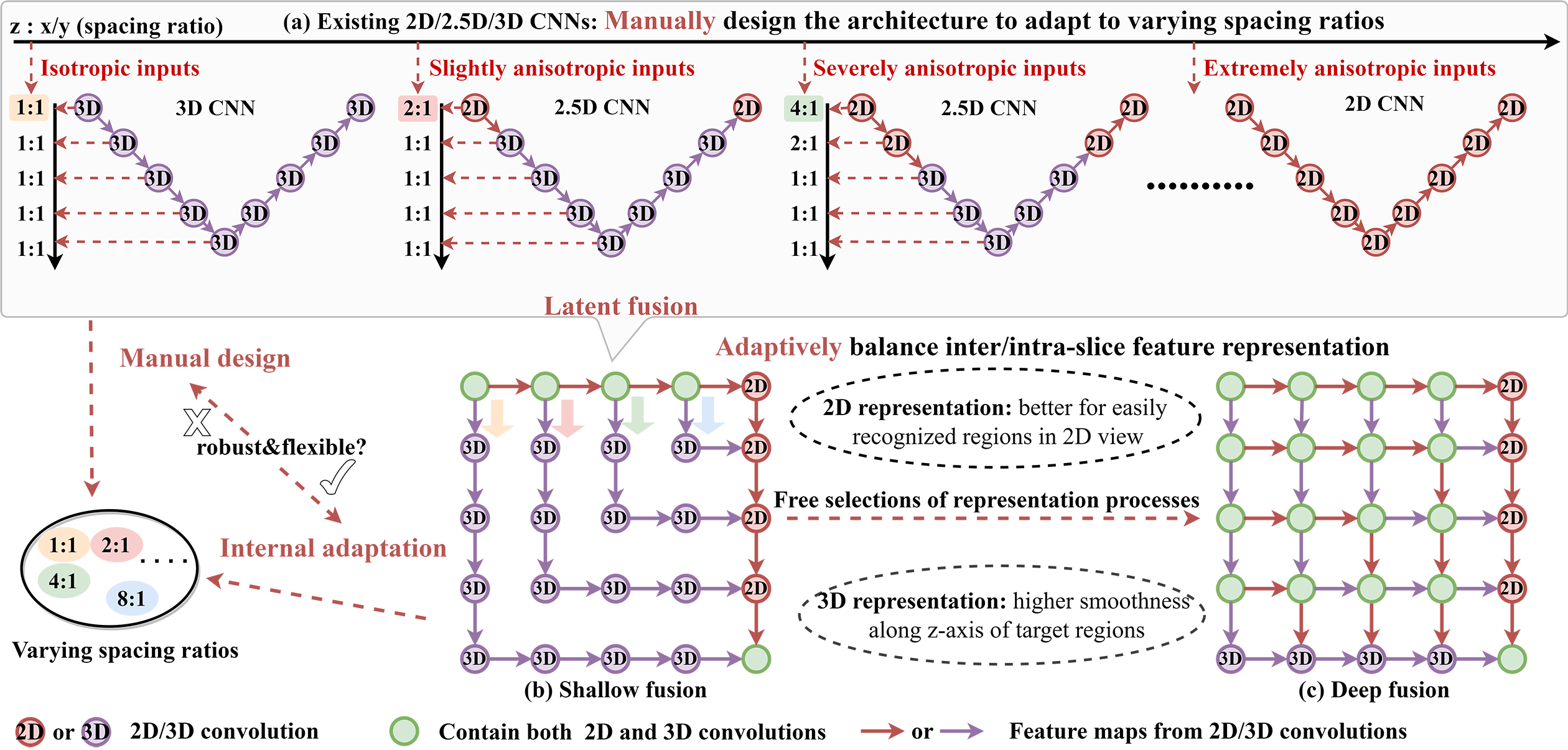}
    \caption{The rethinking of 2D/3D Networks for anisotropic medical image segmentation. (a) Rethinking the 2D/2.5D/3D CNNs. 2D CNNs only represent features in $xy$-plane with even receptive fields, making severe underfitting of inter-slice features. 3D CNNs can capture inter-slice correlations but suffer from the overrepresentation for highly discontinuous long-range sparse information along $z$-axis, which makes the network prone to over-fitting. 2.5D CNNs first use 2D convolutions and pooling for intra-slice feature extraction until the resolution becomes isotropic before performing 3D convolutions, suffering from poor adaptability. (b) We first make a shallow latent fusion of multiple 2D/2.5D/3D CNNs to achieve a multi-path representation process, adapting to the variation of spacing ratios. (c) We further make a deep latent fusion of multi-dimensional convolutions, extending the multi-path network to a mesh architecture, achieving free selections of representation processes via learning, thus realizing balanced representation for anisotropic information.}
    \label{motivation}
\end{figure*}

Due to the limitations of using vanilla 3D or 2D CNNs for the segmentation of anisotropic volumes, some alternatives have been proposed to take the advantages of 3D and 2D convolutions simultaneously: \textbf{1)} 2.5D CNNs \cite{wang2019automatic}. Instead of directly feeding images with anisotropic spacing into 3D CNNs, 2.5D CNNs perform 2D convolution (3$\times$3$\times$1) and pooling (2$\times$2$\times$1) until the spacing of $x/y$-axis is increased to a similar level of $z$-axis before performing 3D convolution, thus roughly achieving balanced representation in 3D field. However, once the spacing ratio of each axis changes, these networks have to be adjusted manually and retrain to adapt (Figure \ref{motivation}(a)). \textbf{2)} Ensemble of 2D and 3D CNNs. 2D CNNs have isotropic receptive fields in $xy$-plane but fail to represent inter-slice features, while 3D CNNs are the opposite. To combine the merits of 2D and 3D CNNs, many methods integrate 2D and 3D CNNs in a serial \cite{lee2015recursive,xia2018bridging} or parallel manner \cite{zheng2019new,isensee2021nnu}. However, imbalanced representation of sparse inter-slice information and dense intra-slice information still exists deeply inside these result-level serial or parallel combinations owing to their independent multi-dimensional representation processes, limiting the final segmentation accuracy.

In this paper, we propose a novel MNet for anisotropic medical image segmentation, which represents sparse inter-slice information and dense intra-slice formation in a balanced way, avoiding the network prone to under/over-fitting to inter-slice information, thus achieving excellent segmentation performance and strong adaptability to the aggravation of anisotropic degree. Specifically:

\paragraph{Innovation 1}: Latent fusion of representation processes. MNet adaptively balances the representation inter axes in the learning process, instead of determining the manner about how to adjust spacing ratio before training. \textbf{1)} We first propose the shallow latent fusion which simultaneously fuses multiple 2D/2.5D/3D CNNs (Figure \ref{motivation}(a)) to achieve a multi-path representation process (Figure \ref{motivation}(b)), thus adapting to the variation of spacing ratios inter axes and achieving preliminarily balanced representation without manual adjustments. \textbf{2)} We further make a deep latent fusion of multi-dimensional convolutions, extending the multi-path network to a mesh architecture (Figure \ref{motivation}(c)), achieving free selections of representation processes via learning, thus realizing balanced representation for sparse inter-slice information and dense intra-slice information. Specifically, at each latent block, if the sparse inter-slice information is over-represented, there is a 2D convolution follows, which doesn't further aggregate inter-slice features. Also, if correlations along $z$-axis are required, the 3D convolution exists as well.

\begin{figure*}[!ht]
    \centering
    \includegraphics[]{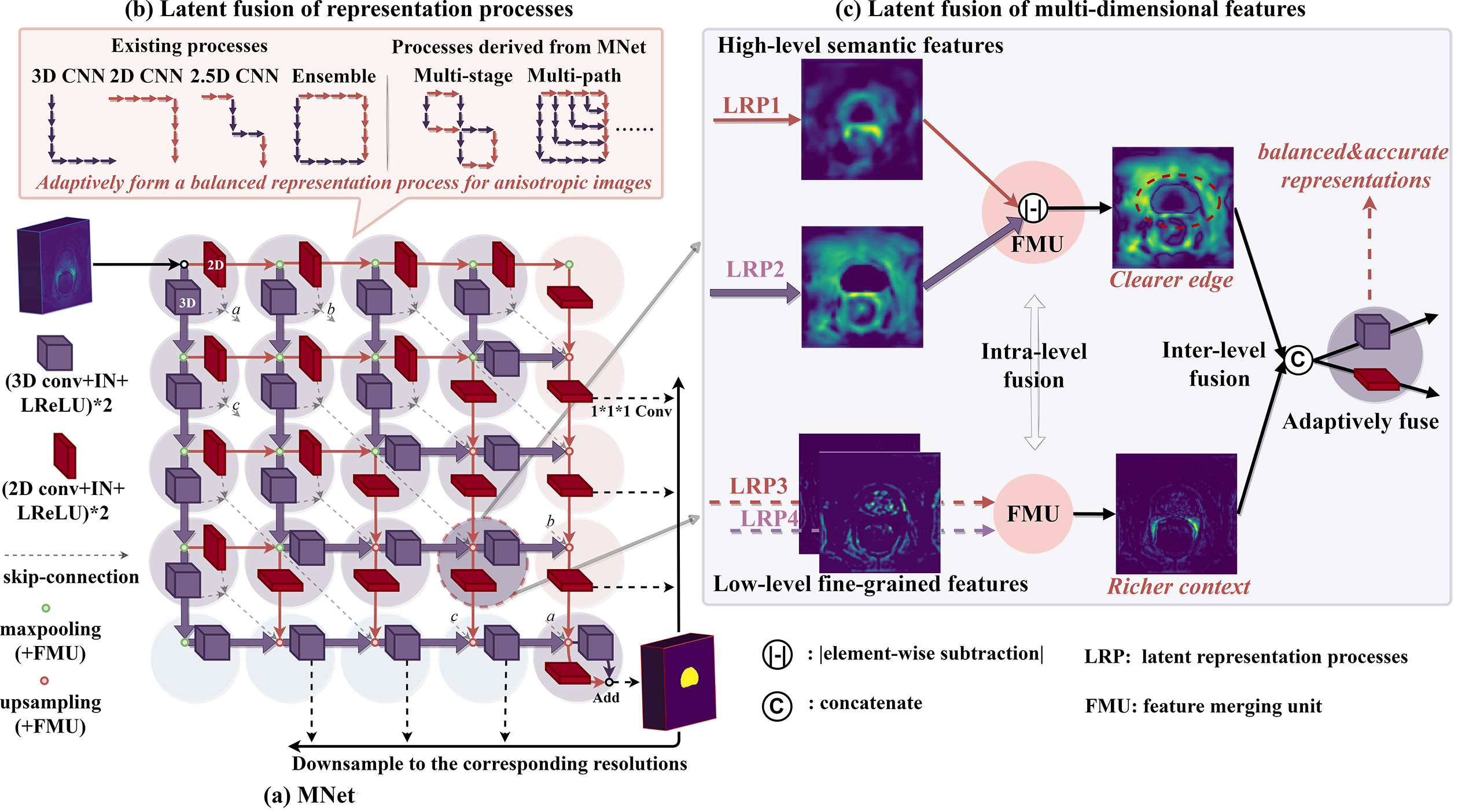}
\caption{
The architecture of our MNet. (a) The mesh structure makes the selections of representation processes unconstrained by embedding multi-dimensional convolutions deeply into latent basic modules. Supervision information is provided to six additional output branches to fully train shallow layers. (b) MNet simultaneously latently fuses plenty of representation processes to adaptively form a balanced representation process via learning for anisotropic information inter axes. (c) MNet latently fuses multi-dimensional and multi-level features inside basic modules, simultaneously taking the advantages of 2D and 3D representations, thus obtaining more accurate modeling for target regions.}
    \label{MNet}
\end{figure*}

\paragraph{Innovation 2}: Latent fusion of multi-dimensional features. MNet contains plenty of latent representation processes (LRP), and the multi-dimensional features from different LRP are fed into basic modules contained in MNet, where the feature-level fusion of anisotropic information is performed for balanced and accurate feature representations. As shown in Figure \ref{MNet}(c), multi-dimensional features are fused to simultaneously take the advantages of 2D and 3D representations that 1) the isotropic receptive field of 2D CNNs leads to the high segmentation accuracy of the easily recognized regions in 2D view, and 2) 3D CNNs pay more attention to the smoothness of 3D organ contour and some structures that are not obvious in 2D view. Therefore, compared with the result-level fusion \cite{zheng2019new,isensee2021nnu}, our feature-level fusion makes better representations from shallow layers to deep layers, obtaining more accurate modeling for target regions.

Our contributions are summarized as follows:

-We propose a novel CNN architecture, MNet, for anisotropic medical image segmentation, which represents sparse inter-slice information and dense intra-slice formation in a balanced way, thus avoiding under- or over-representation to long-range inter-slice features.

-We propose the latent fusion of plenty of representation processes by embedding multi-dimensional convolutions deeply into basic modules contained in MNet, achieving free selections of representation processes via learning, thus balancing representation for sparse inter-slice information and dense intra-slice information adaptively.

-We propose the latent fusion of multi-dimensional features which fuses features from multiple LRP inside each basic module, simultaneously taking the advantages of 2D (high segmentation accuracy of the easily recognized regions in 2D view) and 3D (high smoothness of 3D organ contour) representations, thus obtaining more accurate modeling for target regions.

-We demonstrate the excellent performance of MNet for the segmentation of 3D medical images with anisotropic resolutions by conducting extensive experiments on four widely-used public datasets (CT\&MR).

\section{Related Works}
\subsection{Standalone 2D/3D CNNs}
The representative of the slice-by-slice segmentation methods is 2D U-Net \cite{ronneberger2015u}. The skip connection adopted by U-Net enables the fusing of the fine-grained feature maps from the shallow layer and the coarse-grained but semantic feature maps in the deep layer. Furthermore, UNet++ \cite{zhou2019unet++} adopts a build-in ensemble of U-Nets of varying depths for multi-scale objects and redesigns skip connections for more flexible feature fusion in the decoders. Due to the isotropic voxel spacing of 2D slices ($xy$ plane), the above 2D methods naturally have matched fields of view along $x$- and $y$-axis. However, without constraints from adjacent slices, the 3D organ contours formed by stacking 2D segmentation results are not smooth enough along the $z$-axis.

The segmentation object of 3D CNNs is the 3D volume instead of 2D slices  \cite{milletari2016v}, which enables the extraction of inter-slice information. In the work of \cite{cciccek20163d}, the 2D operation in U-Net is replaced with a 3D counterpart to obtain 3D U-Net, which realizes end-to-end 3D medical image segmentation. To deepen 3D networks without increasing the parameters, ALA-Net \cite{jiang2021ala} adopts the global context encoder (GCE) with the recurrent strategy. To efficiently aggregate long-range information, \cite{wang2020non} propose the non-local U-Nets embedded with the global aggregation block to aggregate long-distance dependencies in biomedical images. The segmentation results of 3D CNNs are smoother than 2D CNNs along the z-axis owing to the utilization of inter-slice information. However, 3D CNNs will make imbalanced feature representation  if the input images are anisotropic.

\subsection{Combination of 2D\&3D CNNs}
Anisotropic spatial resolutions of medical images lead to the uneven physical receptive field in 3D convolutions. To make the receptive field of real world even, the 2.5D U-Net \cite{wang2019automatic} first uses the 2D convolution and pooling for intra-slice feature extraction until the spacing of the $xy$ plane is reduced to a similar level of inter-slice resolution before performing 3D convolution. However, once the resolution ratio of inter axes changes, 2.5D U-Net has to be adjusted manually and retrain to adapt. 

Features of anisotropic images extracted by 2D and 3D CNNs have their own strengths and weaknesses. It is a natural idea to fuse multi-dimensional features for better segmentation performance, and there are already some models that benefit from the fusion\cite{lee2015recursive,xia2018bridging}. In the study of nnU-Net \cite{isensee2021nnu}, the final results are taken from the ensemble of separately trained 2D and 3D U-Nets for better generalization performance. Instead of training 2D and 3D models separately, H-DenseUNet \cite{li2018h} adopts a hybrid feature fusion layer to form better representations with the multi-dimensional features generated by 2D and 3D subnets. The above fusion of 2D and 3D features is mainly for the result-level features, but the independent 2D/3D CNNs still have uneven physical receptive fields inside their feature representation process.

\section{Method}
\subsection{Mesh Architecture}
As shown in Figure \ref{MNet}(a), MNet is composed of 5*5 modules. Modules with purple background contain both the 2D and 3D convolution blocks, while modules with red or blue background have a single 2D or 3D block. Inside the 2D or 3D block are two 3$\times$3$\times$1 or 3$\times$3$\times$3 convolutional layers, both layers are followed by the instance normalization and the LeakyReLU activation function. Using instance normalization instead of batch normalization is mainly because the batch size is limited by GPU memory for 3D images.

Connecting each module with its neighbors, the mesh structure of MNet can be naturally formed. Benefiting from the mesh structure, we have various subnets contained in MNet. The different combinations of 2D and 3D modules make the structure of each subnet different from each other, leading to the extracting of more discriminative features. Encoder-decoder structure is commonly applied in networks for semantic segmentation. Any serial subnet in MNet follows the encoder-decoder structure. Taking the 3D subnet composed of the 3D convolution blocks in the first column and the fifth row as an example, the maxpooling is used for reducing the spatial resolution and aggregating long-range information at the encoding stage of the 3D subnet, which is totally performed four times. Symmetrically, the decoder adopts linear interpolation to gradually recover the spatial resolution of feature maps. The number of filters ($K$) of each convolution block at encoding stage is set to

\begin{equation}
\begin{array}{lr}
     K = 32+16\times (Depth-1),
\end{array}
\end{equation}
where $Depth\in \{1,2,3,4,5\}$, and the filter number of decoder is symmetrical with encoder.

\subsection{Latent Multi-dimensional Feature Fusion}
To combine the merits of 2D and 3D representations, features extracted from different geometric perspectives should be fused. Modules at various depths of MNet are designed to have multiple inputs or/and outputs, so as to make the extraction and fusion of multi-dimensional features throughout the forwarding propagation. The details of the latent fusion in the decoding stage are shown in Figure \ref{MNet}(c). Let $X_{2d}$ and $X_{3d}$ denote the high-level 2D and 3D feature maps from latent representation process $LRP1$ and $LRP2$.  $X_{2d}$ and $X_{3d}$  are first fed into the upsampling layers with scale factors of 1$\times$2$\times$2 ($U_{2d}$) and 2$\times$2$\times$2 ($U_{3d}$), respectively. After that, the multi-dimensional features are passed to the feature merging unit (FMU) which is set to element-wise subtraction followed by the operation of taking the absolute value ($abs$) according to experimental performance, the merged features $F_m$ can be defined as 

\begin{equation}
    F_m = abs(U_{2d}(X_{2d})-U_{3d}(X_{3d})),
\end{equation}
In the process of forwarding propagation, the foreground is gradually detected. However, with the feature level getting higher, the details of the target organ are gradually lost. Therefore, the low-level features with detailed information in the encoder are passed to the decoder to assist in the recovery of the organ contour. Let $X_{2d}^e, X_{3d}^e$ represent the feature maps from  $LRP3$ and $LRP4$, which are passed to the decoding stage through skip-connection. Similarly, the merged features $F_m^e$ can be defined as
\begin{equation}
    F_m^e = abs(X_{2d}^e-X_{3d}^e).
\end{equation}
Subsequently, the channel-wise concatenation is applied to  $F_m^e$ and $F_m$ followed by 2D and 3D convolutions for the extraction of more balanced and accurate features.

\subsection{Latent Representation Process Fusion}
As shown in Figure \ref{MNet}(b), the existing representation processes (e.g., 2D/2.5D/3D CNNs) are latently contained in our MNet, and plenty of novel representation processes can be derived from our MNet, making the selections of representation processes unconstrained, thus balancing representation adaptively. Taking the latent 2.5D CNN as an example, we explain the process about how to balance the representation of anisotropic information.  Assuming that the spacing ratio ($z:x/y$) of the input image is 1:4, the physical distance along $z$-axis between two adjacent voxels is much longer than it along $x/y$-axis in the real world,  so the convolution and maxpooling are first performed only on $xy$-plane two times, gradually increasing the intra-slice spacing, thus making the spacing ratio become 1:1. After that, the isotropic feature maps are fed into the 3D subnet for balanced spatial feature representation. 

\begin{table*}[!ht]
    \centering
    \resizebox{0.95\textwidth}{!}{
    \begin{tabular}{cccccccccc}
         \toprule   
         \multirow{2.5}{*}{\textbf{Methods}}&\multirow{2.5}{*}{\textbf{\makecell[c]{Take anisotropy \\ into account}}} &\multicolumn{2}{c}{\textbf{LiTS}}&\multicolumn{2}{c}{\textbf{KiTS}}&\multicolumn{3}{c}{\textbf{BraTS}}&\textbf{PROMISE}\\
         \cmidrule(r){3-4} \cmidrule(r){5-6} \cmidrule(r){7-9} \cmidrule(r){10-10}
   & &\textbf{Liver} &\textbf{Tumor} &\textbf{Kidney}&\textbf{Tumor} &\textbf{ED}&\textbf{NCR\&NET}&\textbf{ET}&\textbf{Prostate}\\
         \midrule
     3D U-Net&\XSolidBrush &90.1& 51.0& 95.5& 63.7&  81.7& 70.1&82.7&85.6\\
     2D U-Net&\XSolidBrush &91.2&58.1& 95.8& 70.1& 82.6 &71.1&83.5&88.1\\
     \midrule
     2.5D U-Net& \CheckmarkBold & 93.3&58.4&96.2& 79.1& 82.5& 70.5& 83.1&88.4\\
     nnUNet&\CheckmarkBold &94.1&62.0&96.3&79.1&83.3&71.4&\textbf{83.7}&89.5\\
     Our MNet&\CheckmarkBold&\textbf{94.3}&\textbf{66.3}&\textbf{96.4}&\textbf{81.8}&\textbf{83.5}&\textbf{72.0}&83.5&\textbf{89.8}\\
         \bottomrule
    \end{tabular}}
    \caption{Comparison of MNet and other methods on four widely-used datasets in terms of Dice (\%).}
    \label{ResultsOf4Datasets}
    
\end{table*}

\subsection{Deep Optimization for Mesh Architecture}
We perform a deep optimization to ensure the trainable parameters at various depths of mesh architecture are fully optimized. Specially, we adopt the 1$\times$1$\times$1 convolution to form six additional output branches. For the segmentation results of each branch, a resampled label is provided for the loss calculation \cite{9201093,8734700}. The weighted sum of each loss item is taken as the final loss.  The loss function we adopt is the hybrid of cross-entropy loss and dice-coefficient loss, defined as 
 \begin{equation}
    l(X,Y) = -(\frac{2}{C}\sum_{c}\frac{\sum_ix_{ci}y_{ci}}{\sum_ix_{ci}+\sum_iy_{ci}} + \sum_i\sum_cy_{ic}logx_{ic}),
\end{equation}
where $X$ is the prediction with softmax applied and $Y$ is the ground truth. $x/y$ is the voxel contained in $X/Y$. $i$ denotes the index of voxels, while $c\in{C}$ is the class of the current channel. Let $X_{ij}$ denotes the segmentation results of modules in raw $i$ and column $j$, $Y_{ij}$ denotes the corresponding ground truth, the final loss can be defined as
\begin{equation}
 \mathcal{L} = l(X_{55},Y_{55})+\sum_{i=2}^{4}\lambda_i(l(X_{i5},Y_{i5})+l(X_{5i},Y_{5i})),
\end{equation}
where $\lambda=(\frac{1}{2})^{5-i}$ is the weight of each loss item. 

\section{Experiments}
\subsection{Experiments Setup}
\paragraph{Datasets:} Four widely used public datasets, which involve multiple modalities, are selected for comprehensive evaluations. Two CT datasets: \textbf{1)} The Liver and Liver Tumor Segmentation challenge 2017 (LiTS) contains 131 labeled training CT images, target regions are the liver and liver tumors. \textbf{2)} The Kidney and Kidney Tumor Segmentation challenge 2019 (KiTS) provides 210 labeled training CT images whose target regions are the kidneys and kidney tumors. Two MR datasets: \textbf{1)} The Multimodal Brain Tumor Segmentation Challenge 2020 (BraTS) provides 369 labeled training cases. Each case has T1-weighted (T1), post-contrast T1-weighted (T1ce), T2-weighted (T2), and fluid attenuated inversion recovery (Flair) sequences. The target regions are the peritumoral edema (ED), necrotic\&the non-enhancing tumor core (NCR\&NET), and enhancing tumor (ET). \textbf{2)} The T2 MR dataset of the PROMISE challenge 2012 contains 50 labeled training cases, the segmentation target is the prostate.

\paragraph{Implementation Details:} The stochastic gradient descent (SGD) with a momentum of 0.99 is selected as the optimizer. The initial learning rate (0.01) is gradually reduced according to the "poly" learning rate policy \cite{chen2017deeplab}, and the maximum epoch is set to 500. Following the default setting of nnU-Net, the batch sizes for LiTS, KiTS, BraTS, and PROMISE are  2, 2, 4, 2, while the patch sizes are set to 40$\times$224$\times$192, 32$\times$224$\times$224,  28$\times$192 $\times$160,  16$\times$320$\times$320 respectively. Extensive data augmentation techniques are employed to improve the generalization ability (\cite{isensee2021nnu}). Deep supervision and the same model training\&selection schema are applied to all the involved networks for fairness. The experiments are performed with a 32GB V100 GPU. Networks implemented with MindSpore\footnote{https://www.mindspore.cn/} and PyTorch are available at: https://github.com/zfdong-code/MNet.

\paragraph{Evaluation Metric:} To quantitatively evaluate the segmentation performance, the Dice similarity coefficient (Dice) is used as the metric in this work. Dice is the widely used geometrical metric \cite{taha2015metrics}, which measures the overlapping between the prediction and the manual label (higher is better). 

\paragraph{Comparison Settings:}
Extensive experiments are performed on four datasets to fully evaluate the performance of the proposed MNet. Each dataset is randomly split into the training set (80\%) and testing set (20\%). After training for 500 epochs, the model of the final epoch is used for testing. The training processes for all the networks are automatically finished by the nnUNet framework for fairness. \textbf{1) Effectiveness for anisotropic medical images}: We first unify the spacings along $z$-axis of the four datasets to 5mm, which is a common spacing of thick-slice medical images, and their spacings along $x/y$-axis are all below 1mm, then, we compare our MNet and other methods to verify the effectiveness of our innovations. \textbf{2) Adaptability to aggravation of anisotropic degree}: After demonstrating the advanced nature for the segmentation of images with large inter-slice spacing (i.e., 5mm), we further explore the adaptability to the inter-slice spacing changes of MNet and other methods. We unify the spacing along $z$-axis of LiTS dataset five times, obtaining five datasets with inter-slice spacings from 1mm to 5mm to train all the networks. \textbf{3) Selection of feature merging unit}: Finally, we compare the performance of MNet with different feature merging units on the four datasets.

\subsection{Comparison with State-of-the-art Methods}
\subsubsection{Effectiveness for Anisotropic Medical Images} The proposed MNet consistently achieves the best performance on four widely-used datasets compared with other advanced methods. Three valuable conclusions can be drawn from Table \ref{ResultsOf4Datasets}: \textbf{1)} The methods with anisotropy taken into consideration significantly outperform the others. The top segmentation results are rarely produced by generic approaches, which indicates the anisotropic nature does bring challenges to segmentation tasks and the necessity of our innovations. \textbf{2)} The best results are almost always produced by our MNet.  Among methods taking anisotropy into account, 2.5D UNet serially combines 2D and 3D convolutions, making the features fed into 3D convolutions are roughly isotropic, achieving higher performance than standalone 2D and 3D CNNs. The architecture of nnUNet can be dynamically adjusted according to the properties of the given dataset without any manual intervention, resulting in powerful generalization ability, thus outperforming 2.5 UNet. Instead of determining the manner about how to adjust spacing ratio before training, our MNet adaptively balances the representation inter axes in the learning process, owing to its flexible latent extraction and fusion of multi-dimensional representations. \textbf{3)} MNet outperforms other methods by a large margin on fragile structures (e.g., tumors). The results of MNet on tumors of LiTS (66.3) and KiTS (81.8) are more than 10 percent higher than 3D UNet (51.0 and 63.7), demonstrating the discontinuity inter-slice on fragile structures is much more severe than it on large target regions (e.g., kidney), where especially need our balanced representation processes.

\begin{figure}[!t]
    \centering
    \includegraphics[]{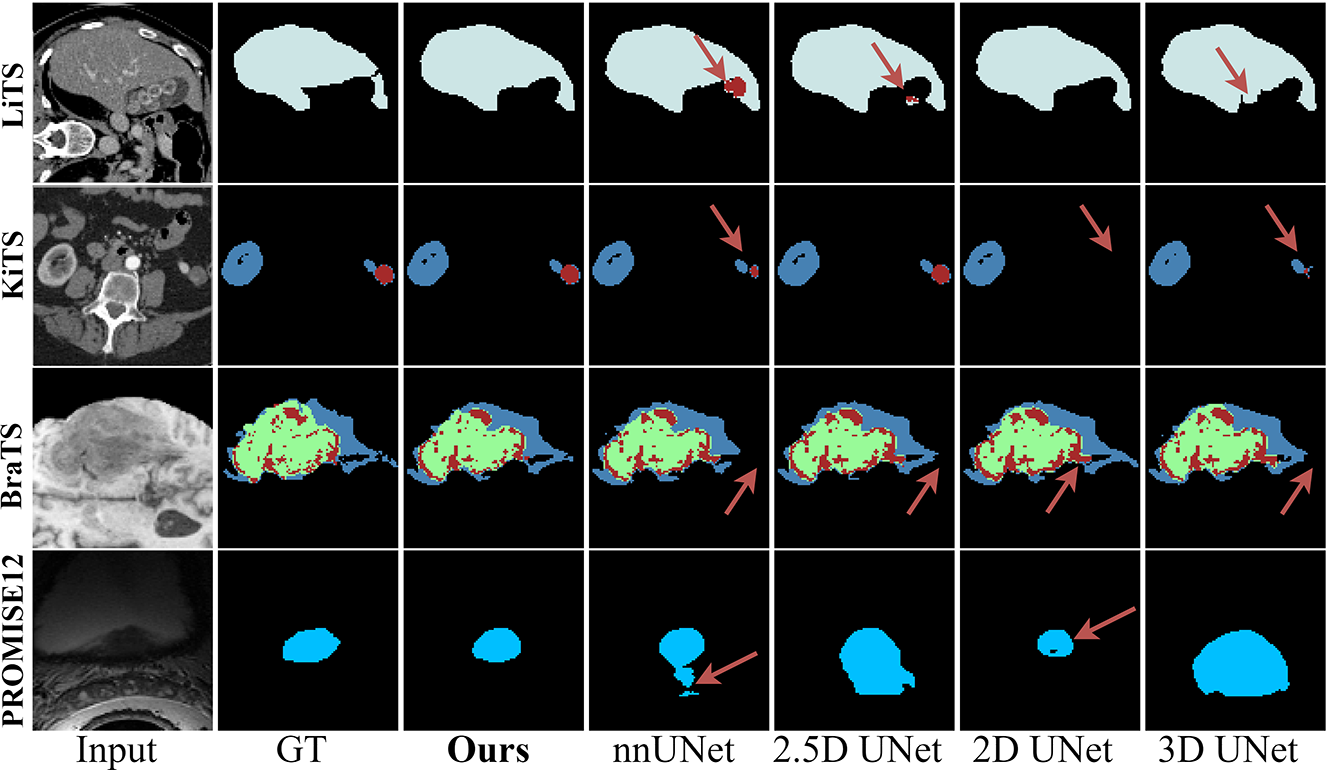}
    \caption{Visualization of segmentation results obtained by different methods for qualitative comparison, which have been cropped for clarity.}
    \label{segComparision}
\end{figure}

To make a qualitative analysis of the segmentation performance, we visualize the segmentation results of the four public datasets. As shown in Figure \ref{segComparision}, the results of MNet are closest to the ground truth (GT) compared with other methods.

\begin{figure}[!b]
    \centering
    \includegraphics[]{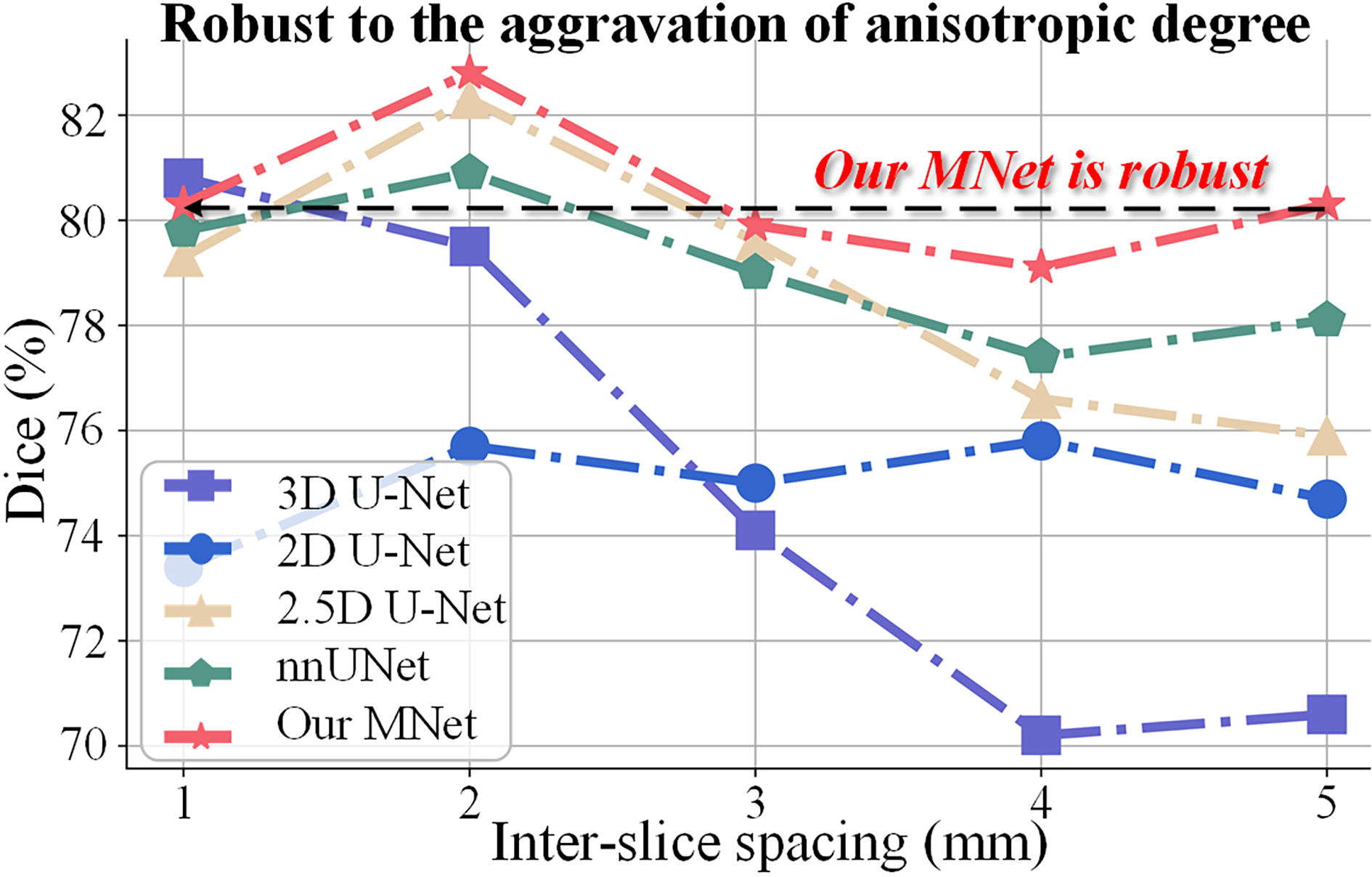}
    \caption{Adaptability to aggravation of anisotropic degree. Our MNet is robust to the aggravation of anisotropic degree, while the performance of other approaches drops drastically. Mean Dice of the liver and tumors are used as evaluation metric.}
    \label{robust}
\end{figure}

\subsubsection{Adaptability to Aggravation of Anisotropic Degree} The proposed MNet has strong adaptability to the aggravation of anisotropic degree in medical images. The spacing along $z$-axis can vary largely from one dataset to another, to explore the adaptability of different approaches to the variation, five datasets derived from the LiTS dataset are separately used to train the methods of comparison. Specifically, to obtain datasets with continuously varying inter-slice spacings, we resample the LiTS dataset five times with five target inter-slice spacings (1mm to 5mm), while the intra-slice spacing keeps the same (0.77mm$\times$0.77mm).

As shown in Figure \ref{robust}, \textbf{1)} our MNet is robust to the aggravation of anisotropic degree, making the mean Dice around 80\%, outperforming other approaches by a large margin (10\% higher than 3D UNet when spacing is 5mm).  \textbf{2)} The performance of 2D UNet is stable but never achieves a high level owing to the lack of information along $z$-axis.  \textbf{3)} 3D UNet performs well when the spacing is roughly isotropic (0.77$\times$0.77$\times1mm^3$), however, with the inter-slice continuity getting lower, fails to balance the representations between $x/y$- and $z$-axis, leading to dramatic performance degradation.  \textbf{4)} Based on the automatic adjustments to the network architecture, nnUNet avoids large performance degradation, but its adaptability is weaker than our MNet since external adjustments made before training cannot well suit the anisotropic nature like our latent multi-dimensional representation balancing learned in the training process.

\subsection{Selection of Feature Merging Unit}
The fusion of features from multiple latent representation processes achieves balanced and accurate representation for anisotropic information. We perform experiments on four datasets to explore the relationship between segmentation performance and different feature merging approaches. Sum makes the merged features contain richer information, while Sub enhances the differences between multiple inputs. As shown in Table \ref{FMU_table}, two types of FMU consistently achieve high and similar performance across the four widely-used datasets, demonstrating what really counts is the behavior of latent fusion and extraction of multi-dimensional representations instead of the specific design of FMU.

\begin{table}[!tbp]
    \centering
    \resizebox{\linewidth}{!}{
    \begin{tabular}{cccccc}
         \toprule
         \textbf{Methods}&\textbf{LiTS} &\textbf{KiTS}&\textbf{BraTS}&\textbf{PROMISE}&\textbf{Mean}\\
         \midrule
         \textbf{Sum}   &79.3& 88.3& \textbf{79.9}&\textbf{90.0}&84.4\\
         
         \textbf{Sub}   & \textbf{80.3}& \textbf{89.1}&79.7&89.8&\textbf{84.7}\\
         \bottomrule
    \end{tabular}}
    \caption{Comparison between different types of feature merging unit (FMU) on the four public datasets in terms of Dice (\%). The reported results are the mean Dice of different target regions. Sum and Sub indicate the summation and absolute value of the subtraction, respectively.}
    \label{FMU_table}
\end{table}

\section{Conclusion}
In this paper, we propose MNet for anisotropic 3D medical image segmentation which can represent sparse inter-slice information and dense intra-slice information in a balanced way, thus avoiding under- or over-representation to inter-slice features. Instead of determining the manner about how to adjust spacing ratio before training, our MNet adaptively balances the representation inter axes in the learning process, owing to its free latent extraction and fusion of multi-dimensional representations. Extensive experiments are performed on four widely-used public datasets, results demonstrate the proposed MNet not only outperforms the methods of comparison but also has outstanding adaptability to the aggravation of anisotropic degree.

\section*{Acknowledgments}
This work was supported in part by the National Key Research and Development Program of China (No. 2021ZD0113202), in part by the National Natural Science Foundation under grants (61828101), CAAI-Huawei MindSpore Open Fund, CANN(Compute Architecture for Neural Networks), Ascend AI Processor, and Big Data Computing Center of Southeast University.








\bibliographystyle{named}
\bibliography{ijcai22}

\end{document}